\documentclass[11pt]{article}
\pdfoutput=1
\usepackage[utf8x]{inputenc}
\usepackage{amssymb,amsmath,mathrsfs,enumerate,appendix}
\usepackage{graphicx,rotate,multicol, multirow}
\usepackage{cite}
\usepackage{braket}
\usepackage[normalem]{ulem}
\usepackage{wrapfig}
\usepackage[textfont=it,labelfont=bf]{caption}
\usepackage{bm}
\usepackage[colorlinks=true,
linkcolor=red,
urlcolor=blue,
citecolor=blue]{hyperref}
\usepackage{lineno}
\usepackage{color}
\usepackage{bbold}
\long\def\rpl#1!!#2!!{\textcolor{red}{#1} \textcolor{blue}{#2}}

\def\@seccntformat#1{\@ifundefined{#1@cntformat}%
	{\csname the#1\endcsname\quad}
	{\csname #1@cntformat\endcsname}
}

\makeatother

\newcommand{\bea}{\begin{eqnarray}}
\newcommand{\eea}{\end{eqnarray}}

\newcommand{\be}{\begin{equation}}
\newcommand{\ee}{\end{equation}}
\def\beq{\begin{equation}}
\def\eeq{\end{equation}}
\newcommand{\ba}{\begin{eqnarray}}
\newcommand{\ea}{\end{eqnarray}}
\def\ifmath#1{\relax\ifmmode #1\else $#1$\fi}

\evensidemargin=\oddsidemargin
\def\Eqn#1{Eq.\ (\ref{#1})}
\def\Eqs#1#2{Eqs.\ (\ref{#1}) and (\ref{#2})}

\allowdisplaybreaks

\title{\Large\bf 
	New physics interpretations for nonstandard values of
	$h\to Z\gamma$
}

\author{
	\sf 
	Rafael Boto$^{a,}$\footnote{rafael.boto@tecnico.ulisboa.pt},
	Dipankar Das$^{b,}$\footnote{d.das@iiti.ac.in},
	Jorge C. Romão$^{a,}$\footnote{jorge.romao@tecnico.ulisboa.pt},
	Ipsita Saha$^{c,}$\footnote{ipsita@iitm.ac.in},
	Joao P. Silva$^{a,}$\footnote{jpsilva@cftp.ist.utl.pt}
	\\[3mm]
	\small\em
	$^a$ Centro de F\'isica Te\'orica de Part\'iculas-CFTP and Departamento de
	F\'isica,  Instituto Superior T\'ecnico,\\  \small\em
	Universidade de Lisboa, Av
	Rovisco Pais, 1, P-1049-001 Lisboa, Portugal \\ 
	\small\em
	$^b$Indian Institute of Technology (Indore), Khandwa Road, Simrol,
	Indore 453 552, India \\
	\small\em
	$^c$Department of Physics, Indian Institute of Technology Madras, Chennai 600036, India
}

\date{}

\begin{document}
	
	
\maketitle
\renewcommand*{\thefootnote}{\arabic{footnote}}
\setcounter{footnote}{0}
	
\begin{abstract}
Current measurement of the $h\to Z\gamma$ signal strength invite us to speculate about
possible new physics interactions that exclusively affect $\mu_{Z\gamma}$ without altering
the other signal strengths. Additional consideration of tree-unitarity enables us to correlate
the nonstandard values of $\mu_{Z\gamma}$ with an upper limit on the scale of new physics.
We find that even when $\mu_{Z\gamma}$ deviates from the SM value by only $20\%$,
the scale of new physics should be well within the reach of the LHC.
\end{abstract}
	
	\maketitle
	
The loop-induced decay modes of the Higgs boson~($h$) have been impactful in many
different aspects of Higgs physics. The decay $h\to \gamma\gamma$, in particular,
played a pivotal role in the discovery of the Higgs boson\cite{ATLAS:2012yve,CMS:2012qbp}.
Such loop-induced Higgs couplings have also been proved useful in sensing the presence of
new physics beyond the Standard Model~(BSM) through new loop
contributions arising from additional nonstandard particles\cite{Bhattacharyya:2014oka}.
This is essentially how the sequential fermionic fourth generation
models fell out of favor\cite{Kribs:2007nz,Eberhardt:2012gv,Djouadi:2012ae}.
These loop-induced Higgs couplings can also provide important insights into
the constructional aspects of the scalar extensions of the SM. Measurements of these
couplings can severely restrict the fraction of nonstandard masses that can be
attributed to the electroweak vacuum expectation value,\cite{Bhattacharyya:2014oka, Bandyopadhyay:2019jzq}
 thereby providing nontrivial information
about the mechanism of electroweak symmetry breaking.

Now that a preliminary measurement of $h\to Z\gamma$ signal strength has become
available, it opens up new avenues to investigate the nature of new physics that may lie
beyond the SM. The currently measured value stands at\cite{ATLAS:2020qcv,CMS:2023mku,CMS:2022ahq}
\begin{eqnarray}
	\mu_{Z\gamma} = 2.2 \pm 0.7 \,,
	\label{e:Zgval}
\end{eqnarray}
which, although not statistically significant yet, may be indicative of an enhancement
compared to the corresponding SM expectation~\cite{Buccioni:2023qnt}, $\mu_{Z\gamma} = 1$. This poses a
rather curious question that if the measurement of $\mu_{Z\gamma}$ settles to a
nonstandard value  while $\mu_{\gamma\gamma}$ is consistent with the SM expectation,
then what kind of new physics would be required to reconcile such observation?
Given the current value of $\mu_{Z\gamma}$, such a possibility might not be
far-off and, from a theoretical standpoint, we must prepare ourselves to
accommodate such an outcome.

\begin{figure}[ht!]
	\centering
	\includegraphics[width=60mm]{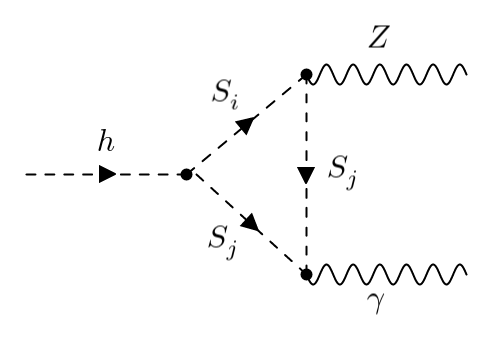}
	\includegraphics[width=60mm]{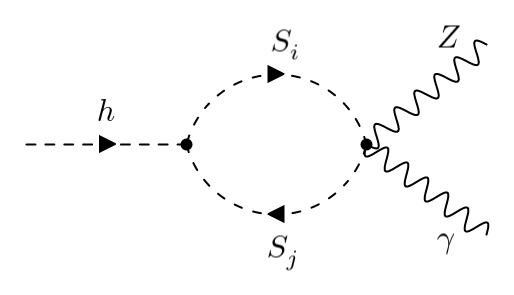}
	\caption{Representative Feynman diagrams that give additional contributions to $\mu_{Z\gamma}$
		exclusively.}
	\label{f:FD}
\end{figure}

\begin{figure}[htbp!]
	\centering
		\includegraphics[width=80mm]{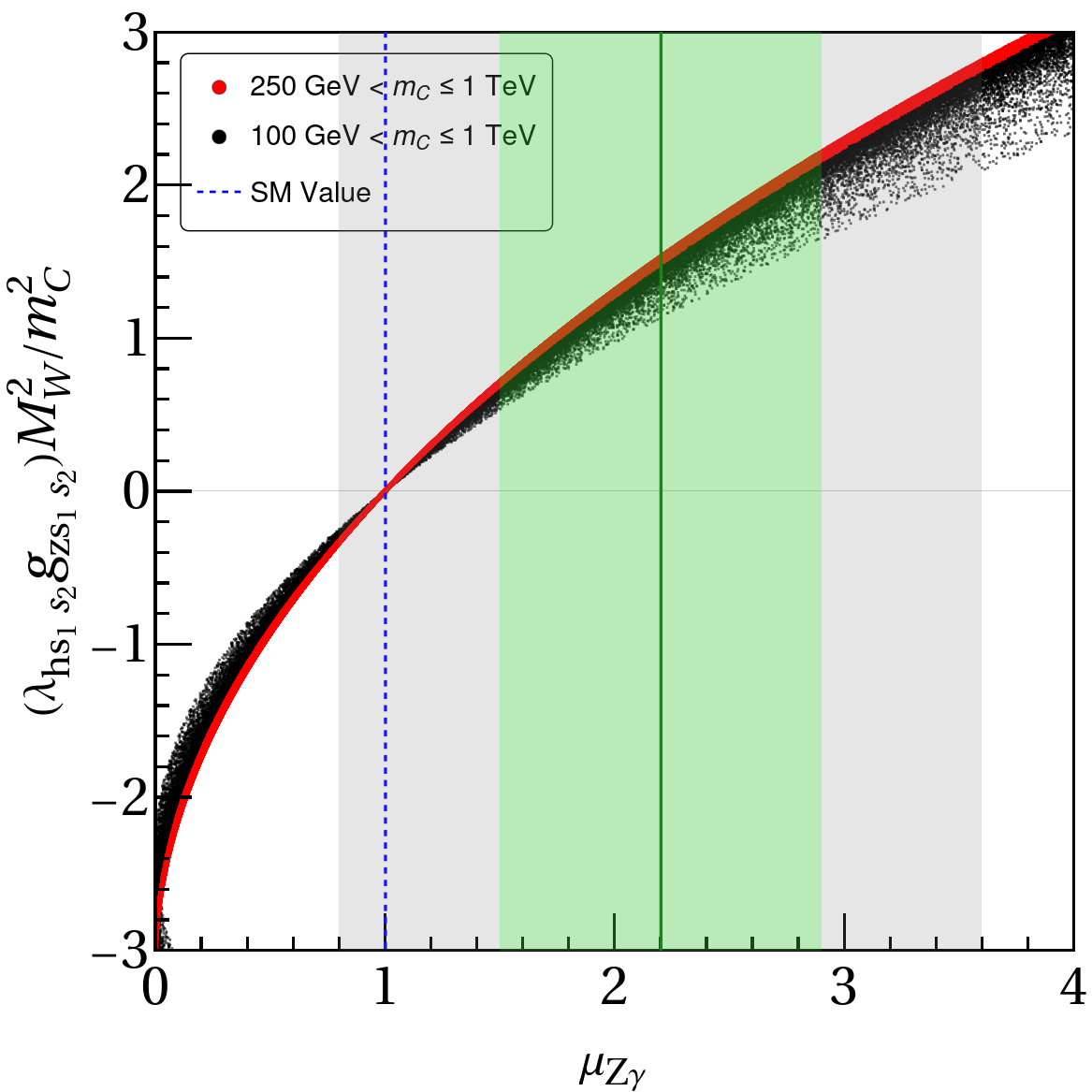}
		\caption{Required values of $f(\mu_{Z\gamma})$ (defined in \Eqn{e:fmu}) as a function of $\mu_{Z\gamma}$.
			The common charged scalar mass ($m_{C_1}=m_{C_2}=m_C$) has been scanned within the range
			[100~GeV, 1~TeV] for the black region and within [250~GeV, 1~TeV] for
			the thin red region.
			The dark-green solid vertical line marks the currently measured central value of $\mu_{Z\gamma}$ and the light-green
			and gray vertical bands around it correspond to the $1\sigma$ and $2\sigma$ ranges respectively. The dashed blue
			vertical line denotes the SM value of $\mu_{Z\gamma}$.
			In making this plot the unitarity conditions of \Eqs{ineq:h}{ineq:dirprod} have been satisfied. 
	}
		\label{f:scalar1}
\end{figure}

%
It is important to realize that, in the usual BSM scenarios, the new physics contributions
affect $\mu_{\gamma\gamma}$ and $\mu_{Z\gamma}$ in a correlated manner\cite{Gunion:1989we,Djouadi:2005gi}. However, if we
are to keep $\mu_{\gamma\gamma}$ intact at its SM value, we must seek new interactions
that exclusively contribute to $\mu_{Z\gamma}$ without altering $\mu_{\gamma\gamma}$.
A little contemplation reveals that `off-diagonal' couplings of the Higgs and the
$Z$-boson would achieve this goal without much hardship. To illustrate this prescription,
let us assume that there exist new charged scalars with couplings parametrized in
the following manner:\footnote{
A similar exercise can also be done assuming the presence of extra charged fermions or vector bosons
possessing analogous off-diagonal couplings.
}
\begin{eqnarray}
\label{e:scalar}
	{\mathscr L}_S^{\rm int} &=& \lambda_{hs_is_j} M_W \, h\, S_i^{+Q}S_j^{-Q} + i\, g_{zs_is_j}
	Z^\mu \left\{\left(\partial_\mu S_i^{+Q}\right)S_j^{-Q} -\left(\partial_\mu S_j^{-Q}\right)S_i^{+Q} \right\} 
	\nonumber \\
	&& + 	{e Q g_{zs_is_j} A^\mu Z_\mu S_i^{+Q}S_j^{-Q}} + g_{zzs_is_j} Z^\mu Z_\mu S_i^{+Q}S_j^{-Q}
	+ {\rm h.c.} \,, 
\end{eqnarray}
where $M_W$ is the $W$-boson mass, $e$ is the electromagnetic coupling constant
 and $S_i^{+Q}$ denotes the $i$-th charged scalar
with electric charge $+Q$. Note that the correlation between the trilinear and
quartic couplings should follow from the underlying gauge theory.\footnote{
A connection between $g_{zs_is_j}$ and $g_{zzs_is_j}$ has been established
in Appendix~\ref{a:appA}.} In \Eqn{e:scalar}
we also assume that the off-diagonal couplings corresponding to $i\ne j$ are
overwhelmingly dominant over the diagonal couplings corresponding to $i=j$,
except for the quartic couplings of the form $ZZSS$. Under
these assumptions only $h\to Z\gamma$ will pick up additional contributions
through the Feynman diagrams shown in Fig.~\ref{f:FD}. These diagrams, quite
obviously, can not contribute to $h\to \gamma\gamma$ as the photon, in its tree-level
couplings, does not change particle species.
The uncommon interactions of \Eqn{e:scalar} 
	will become quintessential if $\mu_{Z\gamma}$ settles to a nonstandard value while
the other signal strengths are compatible with the SM.
%
%
The strengths of the couplings required for accommodating nonstandard values
$\mu_{Z\gamma}$ have been presented in Fig.~\ref{f:scalar1}.\footnote{
The general expression for the $h\to Z\gamma$ amplitude may be found in
Ref.~\cite{Hue:2017cph}. Although for simplicity we chose $Q=1$, for
	the general case the quantity on the vertical axis of Fig.~\ref{f:scalar1}
	will be scaled by a factor of $Q$.
	Additionally, in the rest of the text we choose to focus on the
		scenario with only two species of charged scalars.
}
As can be observed from the figure, the quantity $g_{z s_1 s_2} \lambda_{hs_1s_2}/m_C^2$ can be almost pinned down
uniquely as a function of $\mu_{Z\gamma}$, $\frac{f(\mu_{Z\gamma})}{M_W^2}$, in the limit $m_{C1}=m_{C2}=m_C$ where 
$m_{Ci}$ denotes the mass of the
$i$-th charged scalar. With this spirit we may approximately write
\begin{eqnarray}
\label{e:fmu}
\frac{\lambda_{hs_1s_2}g_{z s_1 s_2}}{m_C^2} \approx \frac{f(\mu_{Z\gamma})}{M_W^2} \,,
\end{eqnarray}
with the understanding that $f(\mu_{Z\gamma})=0$ for $\mu_{Z\gamma}=1$, as can be confirmed using Fig.~\ref{f:scalar1}.
For a particular value of $\mu_{Z\gamma}$, the thickness of the black plot arises because $m_C$ is scanned from
relatively low values, 
within the range 100~GeV $< m_C <$ 1~TeV. The thickness of the plot, for practical purposes,
	becomes negligible once we go beyond $m_C \gtrsim 250$~GeV as can be seen from the thin red overlaid region and
	in this case the equality in \Eqn{e:fmu} becomes more robust.
%

Now that the essential strategy to accommodate a nonstandard $\mu_{Z\gamma}$ has been
laid out, it might be reasonable to ask whether the couplings of Eq.~(\ref{e:scalar}) 
have any additional observable consequences which can potentially falsify such a scenario.
A related study will be to investigate whether the couplings of
Eq.~(\ref{e:scalar}) can arise from a more complete gauge theoretical framework.
It is well-known that the off-diagonal charged scalar couplings can emerge
whenever the physical charged scalars are derived from an admixture of two different
$SU(2)_L\times U(1)_Y$ multiplets. The Zee-type\cite{Zee:1980ai} scalar potential
constitutes a good example
of such a scenario. For the Zee-type set-up, dominant off-diagonal
couplings overpowering the diagonal couplings 
can be achieved when the two charged scalars mix maximally\footnote{
Even in the presence of diagonal couplings one may try to keep $\mu_{\gamma\gamma}$ in the
neighborhood of unity by adjusting ${\cal A}^{\rm NP}_{\gamma\gamma} = -2{\cal A}^{\rm SM}_{\gamma\gamma}$
in the $h\to \gamma\gamma$ amplitude. An example of this with fermionic couplings can be
found in a recent work~\cite{Barducci:2023zml}.
A similar effort within the ambit of left-right symmetry\cite{hong2023decays} leads to modifications of
	$\mu_{Z\gamma}$ and $\mu_{\gamma\gamma}$ in a correlated manner resulting in a limitation
	to the possible enhancement in $\mu_{Z\gamma}$.
}\cite{Florentino:2021ybj}.
However, instead of channeling our efforts to construct a specific model, we can follow
a bottom-up method by exploring the high-energy unitarity behaviors of the tree-level
scattering amplitudes\cite{Gunion:1990kf} involving the couplings of \Eqn{e:scalar}\footnote{
	Our approach in this regard is different from previous studies. For
	example, the unitarity bounds considered in Ref.~\cite{Abu-Ajamieh:2021egq} arise mostly from
	modifications in the tree-level couplings of the Higgs boson with the SM
	particles.
Of course it is well known that the unitarity of the theory will be compromised if such
couplings in the SM are tinkered with\cite{Bhattacharyya:2012tj}. We, on the other hand, do
not touch any of the tree-level SM couplings and the new physics interactions we introduce do not
even affect $\mu_{\gamma\gamma}$.
}. Such an analysis is
known to reveal the compatibility of the set of couplings in \Eqn{e:scalar} with a UV-complete gauge
theory\cite{LlewellynSmith:1973yud, Cornwall:1974km}. If the interactions in \Eqn{e:scalar} necessitate additional dynamics accompanying
them, the scattering amplitudes are expected to possess undesirable energy growths which will
lead to violation of tree-unitarity\cite{Lee:1977eg} at high energies. The energy scale at which
unitarity is violated, can be interpreted as the maximum energy scale before which the effects
of new physics must set in to restore unitarity. Such an exercise provides an alternative
strategy to discover the need for additional effects that should be accompanied by
an enhanced $\mu_{Z\gamma}$. As we will show in the appendix, inclusion of proper quartic interaction
of the form $ZZSS$ will neutralize the bad high-energy behaviors.

To demonstrate this explicitly,
we now concentrate on the impact of the $ZZSS$ quartic couplings in \Eqn{e:scalar}, namely,
\begin{eqnarray}
\label{e:zzss}
{\mathscr L}_{ZZSS}^{\rm int} = g_{zzs_is_j} Z^\mu Z_\mu S_i^{+Q}S_j^{-Q}
+ {\rm h.c.} \,. 
\end{eqnarray}
As mentioned earlier, these couplings will be required
to complement the underlying gauge structure. As we show in Appendix~\ref{a:appA},
even in the limit when the rest of the couplings of \Eqn{e:scalar} are purely off-diagonal,
the quartic interactions of \Eqn{e:zzss} should be diagonal with a specific
relation between $g_{zzs_is_j}$ and $g_{zs_is_j}$\footnote{For off-diagonal
couplings in \Eqn{e:scalar} and diagonal couplings in \Eqn{e:zzss} there will
be no additional loop-induced effects for the $hZZ$ vertex as well\cite{Hernandez-Juarez:2023dor} as long as we work with only
	two flavors of charged scalars.}.
With this information, we can now proceed to calculate the amplitude for the process
$Z_L Z_L \to S_1^+ S_1^-$ where the subscript `$L$' represents longitudinal polarization.
In the high-energy limit, $E_{CM}\gg M$, meaning the CM energy is much larger than all
the masses in our current theory, we obtain
\begin{eqnarray}
{\cal M}_{Z_L Z_L \to S_1^+ S_1^-} \approx \frac{2 g_{z s_1 s_2}^2}{M_Z^2}
\left(m_{C_1}^2 -m_{C_2}^2 \right)
 +{\cal O}\left(\frac{M^2}{E_{CM}^2}\right) \,.
 \label{eq:matproc1}
\end{eqnarray}
Thus it is clear that the splitting between the two charged scalar masses is
constrained from unitarity as 
\begin{eqnarray}
\left|\frac{2 g_{z s_1 s_2}^2}{M_Z^2}
\left(m_{C_1}^2 -m_{C_2}^2\right)\right| < 16 \pi\,.
\label{e:split}
\end{eqnarray}
Therefore our simplified assumption of $m_{C_1}=m_{C_2}=m_C$ is manifestly
consistent with the unitarity requirements irrespective of the magnitude of
$g_{z s_1 s_2}$.
%
Next we consider the process $Z_L Z_L \to S_1^+ S_2^-$.
In the high energy limit, the amplitude is found to be
\begin{eqnarray}
	{\cal M}_{Z_L Z_L \to S_1^+ S_2^-} \approx -\frac{g}{2} \lambda_{hs_1s_2} + {\cal O}\left(\frac{M^2}{E_{\rm CM}^2}\right) \,,
	\label{eq:mth}
\end{eqnarray}
where $g$ is the $SU(2)_L$ gauge coupling.
This puts an upper limit on $\lambda_{hs_1s_2}$ as follows
\begin{eqnarray}
	\left|\frac{g}{2} \lambda_{hs_1s_2}\right| < 16 \pi\,.
	\label{ineq:h}
\end{eqnarray}
Finally we note that a direct upper bound on the charged scalar masses can be
placed by considering the scattering process $Z_L S^+_1 \to h S_1^+$. In the high-energy limit the tree-level amplitude can be written as
\begin{eqnarray}
{\cal M}_{Z_L S^+_1 \to h S_1^+} \approx - g_{zs_1s_2} \lambda_{hs_1s_2} \frac{M_W}{M_Z} + {\cal O}\left(\frac{M^2}{E_{\rm CM}^2}\right)\,. 
\label{eq:matproc3}
\end{eqnarray}
Therefore the unitarity constraint should imply
\begin{eqnarray}
\left|g_{z s_1s_2} \lambda_{hs_1s_2}\right| \frac{M_W}{M_Z} < 16\pi \,.
\label{ineq:dirprod}
\end{eqnarray}
This is where the experimental determination of
$\mu_{Z\gamma}$ becomes relevant. A nonstandard value of $\mu_{Z\gamma}$
exclusively, will necessitate such couplings whose strength can be estimated
using \Eqn{e:fmu} as follows
\begin{eqnarray}
	\lambda_{hs_1s_2}g_{z s_1 s_2} = \frac{f(\mu_{Z\gamma})\, m_C^2}{M_W^2} \,.
\end{eqnarray}
Plugging this into \Eqn{ineq:dirprod} we may infer
\begin{eqnarray}
m_C < \sqrt{16 \pi \frac{ M_Z M_W}{\left|f(\mu_{Z\gamma})\right|}} \,.
\label{e:mclim}
\end{eqnarray}
\begin{figure}[htbp!]
	\centering
	\includegraphics[width=0.5\textwidth]{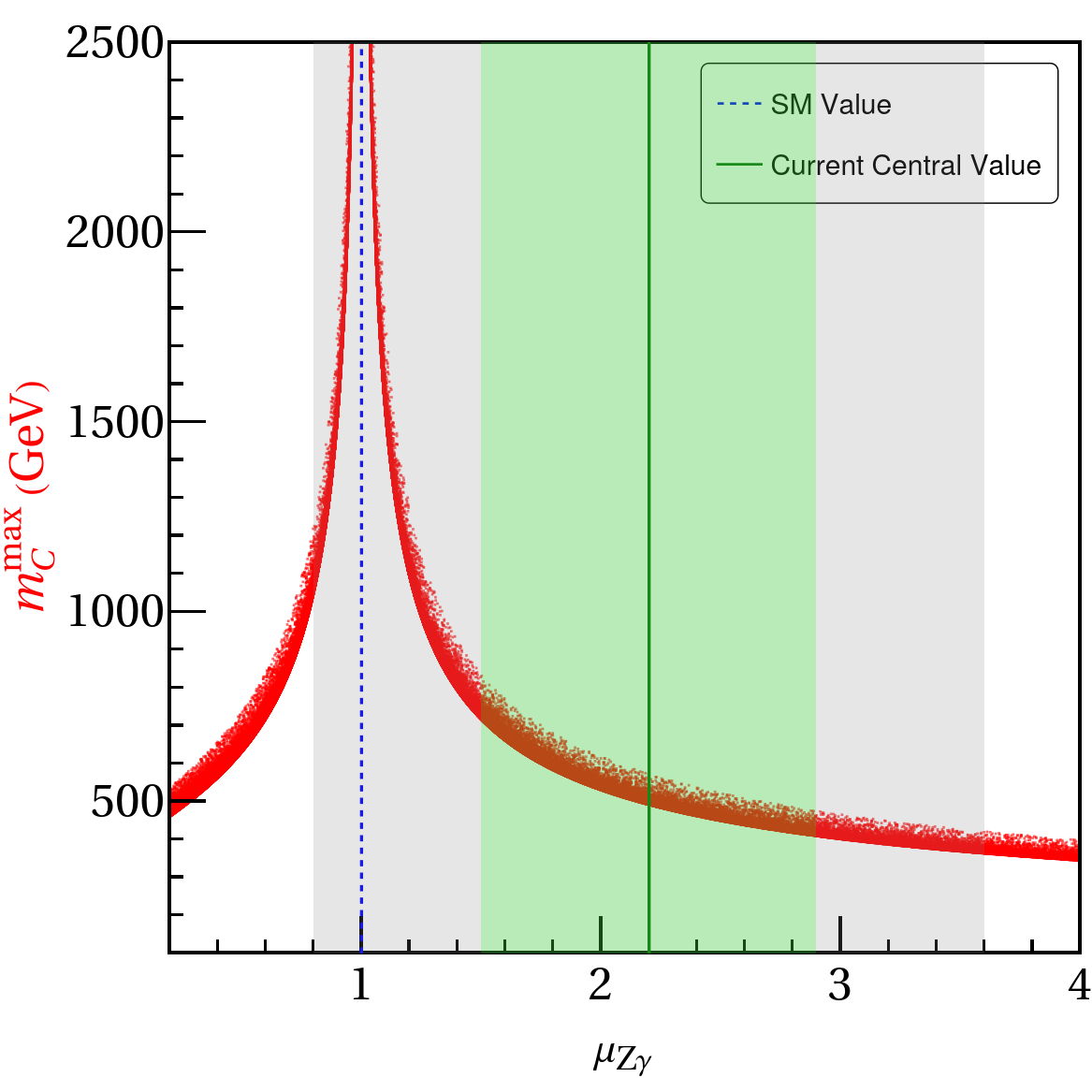}
	\caption{\it The upper limits in \Eqn{e:mclim} plotted against $\mu_{Z\gamma}$
		as the red region respectively.
		The dark-green solid vertical line marks the currently measured central value of $\mu_{Z\gamma}$ and the light-green
		and gray vertical bands around it correspond to the $1\sigma$ and $2\sigma$ ranges respectively. The dashed blue
		vertical line denotes the SM value of $\mu_{Z\gamma}$.
	}
	\label{f:uniNP}
\end{figure}
It should be noted that, as $f(\mu_{Z\gamma}=1)=0$, the upper limit on $m_C$ can be
infinitely large for $\mu_{Z\gamma}=1$, implying that the new physics effects can be safely decoupled in
the SM-limit as expected. However a more intriguing thing to note will be the fact that any
deviation of $\mu_{Z\gamma}$ from the SM value will mandate the intervention of new physics. To quantify
the required proximity of the new physics scale, we plot the right hand side of \Eqn{e:mclim} as the
red region in Fig.~\ref{f:uniNP} where
the value of the function $f(\mu_{Z\gamma})$ is mapped from Fig.~\ref{f:scalar1}.
From Fig.~\ref{f:uniNP} we can see that new physics effects in the sub-TeV regime will be imminent
even when $\mu_{Z\gamma}$ deviates from unity by only $20\%$. In fact, the current central
value of $\mu_{Z\gamma}=2.2$ (marked by the dark-green vertical solid line)
decrees the common charged scalar mass to be below 500~GeV which should be well within
the reach of the LHC.


To summarize, recent experimental data on $\mu_{Z\gamma}$ instigates us to contemplate the possibility
of having all the Higgs signal strengths in excellent agreement with the corresponding SM expectations,
except $\mu_{Z\gamma}$ which deviates substantially from its SM value. Our current article can be 
considered as a theoretical preparation for such an eventuality in a bottom-up manner. We 
provided a general template for the new physics interactions which will exclusively affect $h \to Z \gamma$.
We particularized our strategy with new charged scalars endowed with dominant off-diagonal couplings
with the Higgs and the $Z$ bosons, as exemplified through Eq.~(\ref{e:scalar}). 
However, as we have explicitly shown,
such interactions do not compromise unitarity at high-energies
indicating compatibility with spontaneously broken gauge theories. 
The unitarity constraints have been used to place upper bounds on
the magnitudes of the new couplings. We have then translated them
into an upper
bound on the common charged scalar mass, which can be as low as 500 GeV for the current central value of
$\mu_{Z\gamma}$. 
This means that, if the current non-standard value of $\mu_{Z\gamma}$ becomes statistically
significant as more data accumulates, discovery of new physics effects at the LHC should be just around the
corner. The charged scalars, owing to their couplings with the photon, can be pair-produced by the
Drell-Yan mechanism~\cite{Alves:2005kr,Banerjee:2022xmu}. The subsequent decay modes of these charged scalars will, of course, 
depend on the finer details of the BSM scenario from which they arise. Even if the charged scalars
are stable, they can be probed in the ongoing searches for long-lived charged particles~\cite{Heinrich:2018pkj,ATLAS:2019gqq}. 
This anticipatory experimental scenario may be compared with the status of the LHC in the pre-Higgs-discovery era, as
a win-win machine
in the sense that the LHC should either observe the Higgs-boson or something equivalent, or violation of unitarity at high energies. In a similar spirit, LHC can again act as a win-win experiment for BSM searches if $\mu_{Z\gamma}$ eventually settles towards a nonstandard value. That would definitely be an exciting future to look forward to.

\paragraph{Acknowledgements:}
%
We sincerely thank the anonymous referee for some very useful comments which led
to a substantial improvement of our manuscript.
DD thanks the Science and Engineering Research Board, India for financial support
through grant no. CRG/2022/000565.
IS acknowledges the support from project number RF/23-24/1964/PH/NFIG/009073 and from DST-INSPIRE, India, under grant no. IFA21-PH272.
The work of R.B. is supported in part by the Portuguese
Funda\c{c}\~{a}o para a Ci\^{e}ncia e Tecnologia\/ (FCT)
under contract PRT/BD/152268/2021.
The work of R.B, J.C.R., and J.P.S. is supported in part by FCT under
Contracts CERN/FIS-PAR/0002/2021,
UIDB/00777/2020, and UIDP/00777/2020; these projects are partially funded through POCTI (FEDER),
COMPETE, QREN, and the EU.
\appendix
\setcounter{equation}{0}
\renewcommand{\theequation}{\thesection.\arabic{equation}}
\section{Theoretical conditions for off-diagonal $Z$-boson couplings}
\label{a:appA}
%
%
Let $H_1^{+Q}$ and $H_2^{+Q}$ be two (unphysical) charged scalars originating
from two different $SU(2)_L$ multiplets and therefore having well defined
$T_3$ eigenvalues denoted by $T_3^{(1)}$ and $T_3^{(2)}$ respectively.
Thus, their interactions with the $Z$-boson may be parametrized as
\begin{eqnarray}
	\label{e:quartic}
	{\mathscr L}_H^{Z} &=&  i R_{z_1} \left \{ \left(\partial_\mu H_1^{+Q}\right)H_1^{-Q} - \left(\partial_\mu H_1^{-Q}\right)H_1^{+Q}\right\}Z^\mu + R_{z_1}^2 (H_1^{+Q} H_1^{-Q})(Z^\mu Z_\mu)  
	\nonumber \\
	&& + i R_{z_2} \left \{ \left(\partial_\mu H_2^{+Q}\right)H_2^{-Q} - \left(\partial_\mu H_2^{-Q}\right)H_2^{+Q}\right\}Z^\mu + R_{z_2}^2 (H_2^{+Q} H_2^{-Q})(Z^\mu Z_\mu) 
	 \,, 
\end{eqnarray}
where,
\begin{eqnarray}
	R_{z_i} = \frac{g}{c_w}\left(T_3^{(i)} - Qs_w^2\right) \qquad i=1,2 \,.
\end{eqnarray}
Here $g$ denotes the $SU(2)_L$ gauge coupling, $s_w$ and $c_w$ are the sine and the cosine of the weak mixing angle,
respectively. The physical charged scalars, $S_1^{+Q}$ and $S_2^{+Q}$, should
be obtained by the following rotation:   
\begin{eqnarray}
	\label{e:basisrot}
	\begin{pmatrix}
		H_1^{+Q} \\
		H_2^{+Q} 
	\end{pmatrix}
	&=& \begin{pmatrix}
		\cos \zeta & \sin \zeta  \\
		-\sin \zeta & \cos \zeta \end{pmatrix}
	\begin{pmatrix}
		S_1^{+Q} \\
		S_2^{+Q} 
	\end{pmatrix}
\end{eqnarray}
Substituting Eq.~{\ref{e:basisrot}} into Eq.~{\ref{e:quartic}}, we find,
\begin{eqnarray}
	{\mathscr L}_S^{Z} &=&  i R_{z_1}  \left[  \cos^2 \zeta \left \{  \left(\partial_\mu S_1^{+Q}\right)S_1^{-Q} - \left(\partial_\mu S_1^{-Q}\right)S_1^{+Q}\right\} + \sin^2 \zeta \left \{  \left(\partial_\mu S_2^{+Q}\right)S_2^{-Q} - \left(\partial_\mu S_2^{-Q}\right)S_2^{+Q}\right\} \right.
	\nonumber \\
 && \left. + \sin \zeta \cos \zeta \left \{  \left(\partial_\mu S_1^{+Q}\right)S_2^{-Q} - \left(\partial_\mu S_1^{-Q}\right)S_2^{+Q} + \left(\partial_\mu S_2^{+Q}\right)S_1^{-Q} - \left(\partial_\mu S_2^{-Q}\right)S_1^{+Q}\right\} \right]Z^\mu \nonumber \\
	&& +  R_{z_1}^2 \left[ \cos^2 \zeta \, S_1^{+Q} S_1^{-Q} + \sin^2 \zeta \,  S_2^{+Q} S_2^{-Q} + \cos \zeta \sin \zeta \,  \left\{S_1^{+Q} S_2^{-Q} + S_2^{+Q} S_1^{-Q} \right\}\right](Z^\mu Z_\mu)  	\nonumber \\
	&&  + i R_{z_2}  \left[ \sin^2 \zeta  \left \{  \left(\partial_\mu S_1^{+Q}\right)S_1^{-Q} - \left(\partial_\mu S_1^{-Q}\right)S_1^{+Q}\right\} + \cos^2 \zeta \left \{  \left(\partial_\mu S_2^{+Q}\right)S_2^{-Q} - \left(\partial_\mu S_2^{-Q}\right)S_2^{+Q}\right\} \right.
	\nonumber \\
 && \left. - \sin \zeta \cos \zeta  \left \{  \left(\partial_\mu S_1^{+Q}\right)S_2^{-Q} - \left(\partial_\mu S_1^{-Q}\right)S_2^{+Q} + \left(\partial_\mu S_2^{+Q}\right)S_1^{-Q} - \left(\partial_\mu S_2^{-Q}\right)S_1^{+Q}\right\} \right]Z^\mu \nonumber \\
	&& +  R_{z_2}^2 \left[ \sin^2 \zeta \, S_1^{+Q} S_1^{-Q} + \cos^2 \zeta \,  S_2^{+Q} S_2^{-Q} - \cos \zeta \sin \zeta \,  \left\{S_1^{+Q} S_2^{-Q} + S_2^{+Q} S_1^{-Q} \right\}\right](Z^\mu Z_\mu)  	
	 \,, 
\end{eqnarray}
such that the diagonal couplings are given by,
\begin{eqnarray}
g_{zs_1s_1}&=&\cos^2 \zeta \, R_{z_1}+\sin^2 \zeta \, R_{z_2} = R_{z_1}- \sin^2 \zeta \, \left( R_{z_1} - R_{z_2} \right)   \,, \nonumber\\
g_{zzs_1s_1}&=&\cos^2 \zeta \, R^2_{z_1}+\sin^2 \zeta \, R^2_{z_2} \,, \nonumber\\
g_{zs_2s_2}&=&\sin^2 \zeta \, R_{z_1}+\cos^2 \zeta \, R_{z_2} = \sin^2 \zeta \, \left( R_{z_1} - R_{z_2} \right)  +  R_{z_2}  \,, \nonumber\\
g_{zzs_2s_2}&=&\sin^2 \zeta \, R^2_{z_1}+\cos^2 \zeta \, R^2_{z_2} 
 \,, 
\end{eqnarray}
and off-diagonal,
\begin{eqnarray}
g_{zs_1s_2}&=&\tfrac{1}{2} \left(   R_{z_1}- R_{z_2} \right) \sin 2\zeta  \,, \nonumber\\
 g_{zzs_1s_2}&=&\tfrac{1}{2} \left(   R^2_{z_1}- R^2_{z_2} \right) \sin 2\zeta = g_{zs_1s_2} \left( R_{z_1} + R_{z_2}\right)
 \,, 
\end{eqnarray}
Both trilinear diagonal couplings, $g_{zs_1s_1}$ and $g_{zs_2s_2}$, will
vanish simultaneously if the following conditions are satisfied
\begin{subequations}
	\label{e:offconds}
	\begin{eqnarray}
	&& R_{z_1}=-R_{z_2} \,, \quad \Rightarrow \, T_3^{(1)} + T_3^{(2)} = 2Q s_w^2 \,,
	\label{e:offcond1} \\
	{\rm and,} && \sin \zeta = \frac{1}{\sqrt{2}} \,.
	\label{e:offcond2}
	\end{eqnarray}
\end{subequations}
Under these conditions the remaining couplings take the form,
\begin{subequations}
	\label{e:othercoup}
	\begin{eqnarray}
	g_{zzs_1s_2}  &=&  0  \,, \\
	g_{zs_1s_2} &=& R_{z_1} \,,  \\
	g_{zzs_1s_1}  &=&   R_{z_1}^2 = g^2_{zs_1s_2}  \,, \\
	g_{zzs_2s_2}  &=&   R_{z_1}^2 = g^2_{zs_1s_2} \,.
	\end{eqnarray}
\end{subequations}
Because of the numerical value~\cite{Workman:2022ynf} of $s_w^2 \approx 0.23$, \Eqn{e:offcond1} can 
be approximately satisfied when two singly charged scalars ($Q=1$) arise from
a mixing between an $SU(2)_L$ singlet ($T_3^{(1)}=0$) and an 
$SU(2)_L$ doublet ($T_3^{(2)}=1/2$). The Zee-type model\cite{Zee:1980ai} constitutes a prototypical
example of such a scenario. 
We have verified the existence of allowed points in the parameter space of such a model\cite{Florentino:2021ybj}, which conform to maximal mixing as in
\Eqn{e:offcond2}\footnote{The angle $\zeta$  corresponds to $\gamma$ in the Zee-type model\cite{Zee:1980ai}. } and lead to very suppressed trilinear diagonal couplings with the Higgs and the $Z$-boson as compared to the corresponding off-diagonal couplings. 
Furthermore, the Zee-type model admits
an ‘alignment limit’ which ensures that the lightest CP-even scalar mimics an
SM-like Higgs boson. Therefore, one can achieve the SM-like $hXX$ couplings 
($X$ denotes a massive SM-particle)
by staying in the proximity of `alignment limit' while independently
realizing the maximal mixing ($\zeta \approx 45^\circ$) between the charged scalars
by adjusting the parameters in the scalar potential. 
%

\section{Explicit calculations of the scattering amplitudes}
\label{s:appB}
\begin{figure}[htbp!]
	\centering
	\includegraphics[scale=0.15]{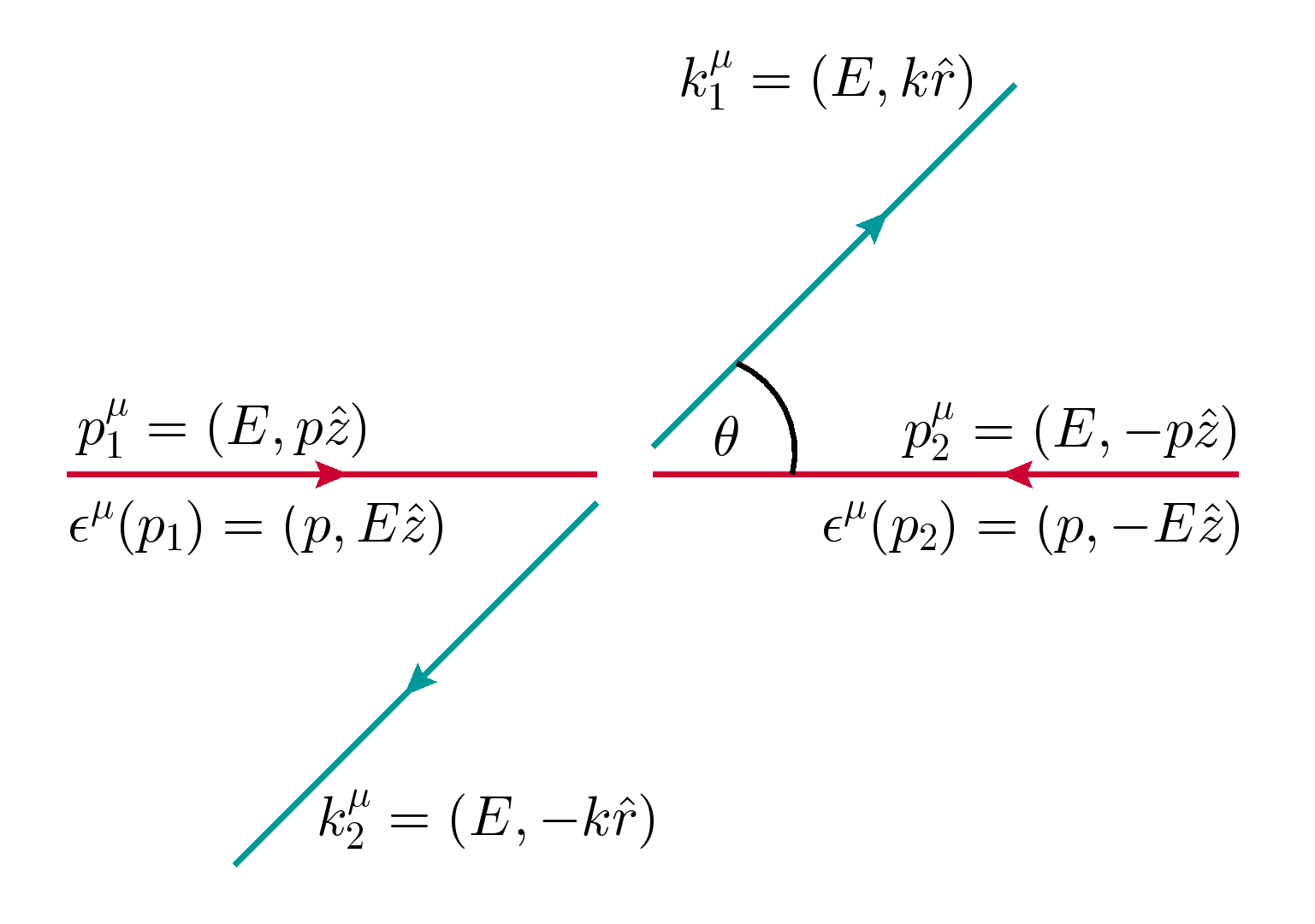}
	\caption{Kinematics in the CM frame for the process in \Eqn{e:P1}.}
	\label{f:kine}
\end{figure}

In this appendix, we show the explicit calculations of the scattering amplitudes discussed in the
main text. First we consider the process
\begin{eqnarray}
	\centering
	\Huge
	Z_L(p_1) + Z_L(p_2) \to S_1^+(k_1) + S_1^-(k_2) \,.
	\label{e:P1}
\end{eqnarray}
%
\begin{figure}[ht!]
	\centering
	\includegraphics[scale=0.17]{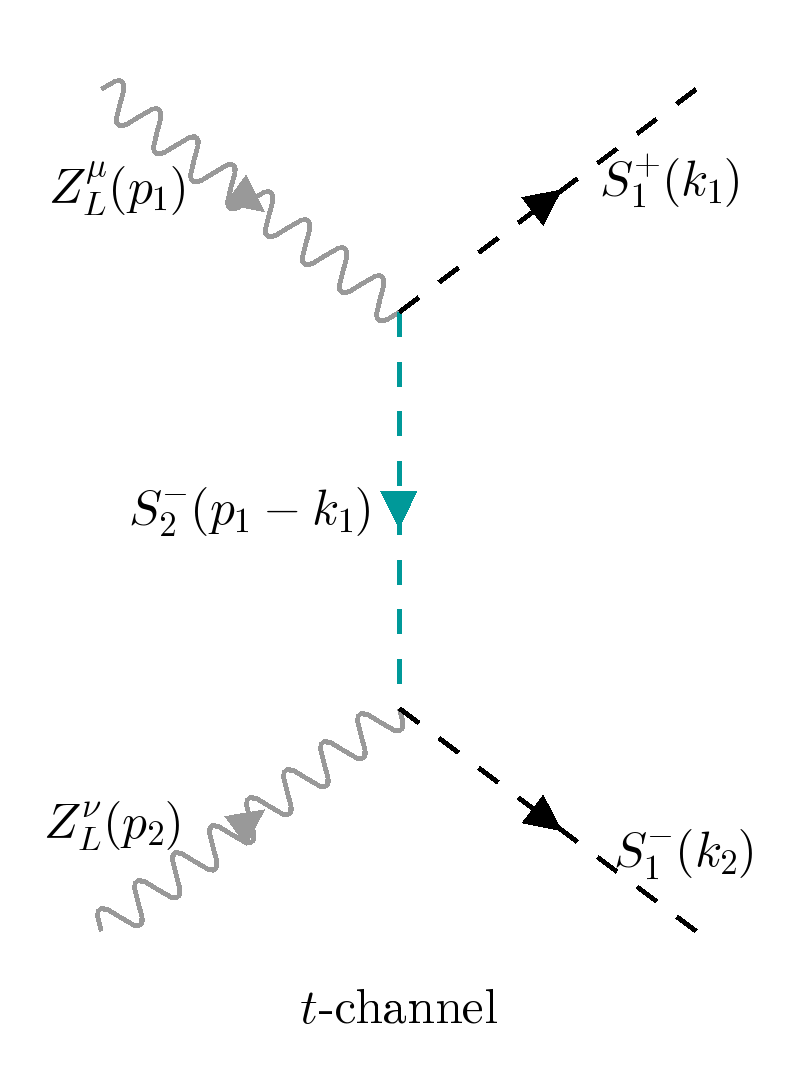}
	\includegraphics[scale=0.17]{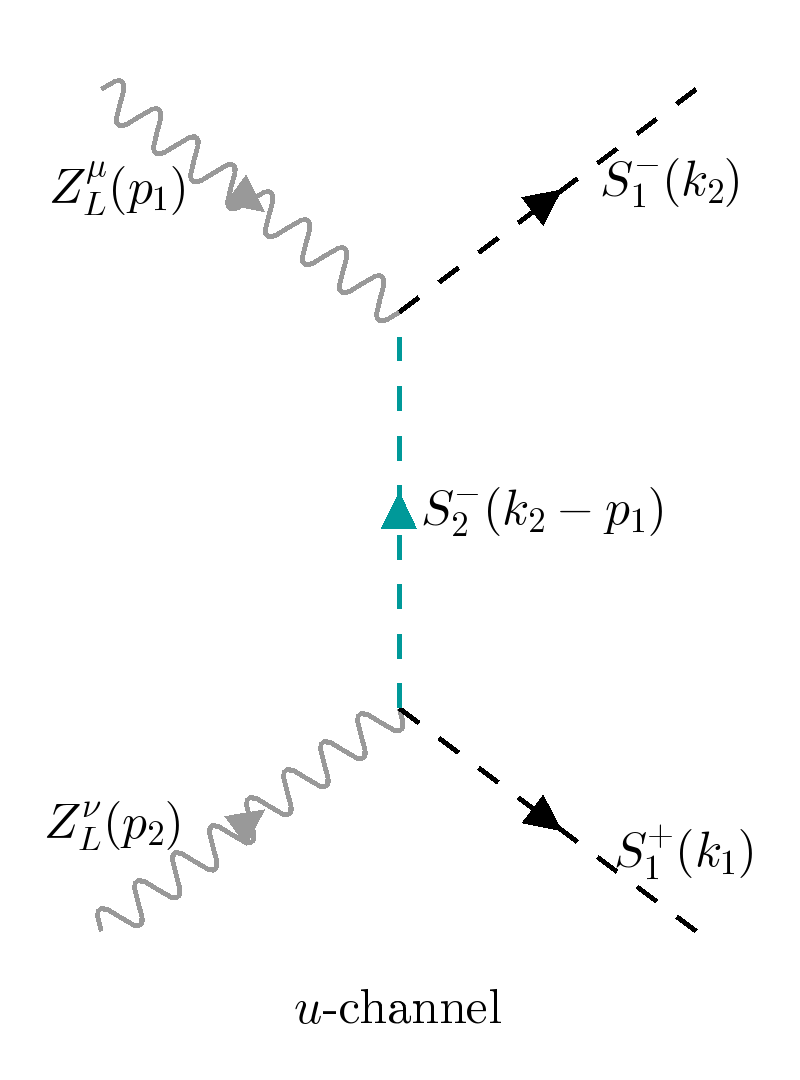}
		\includegraphics[scale=0.15]{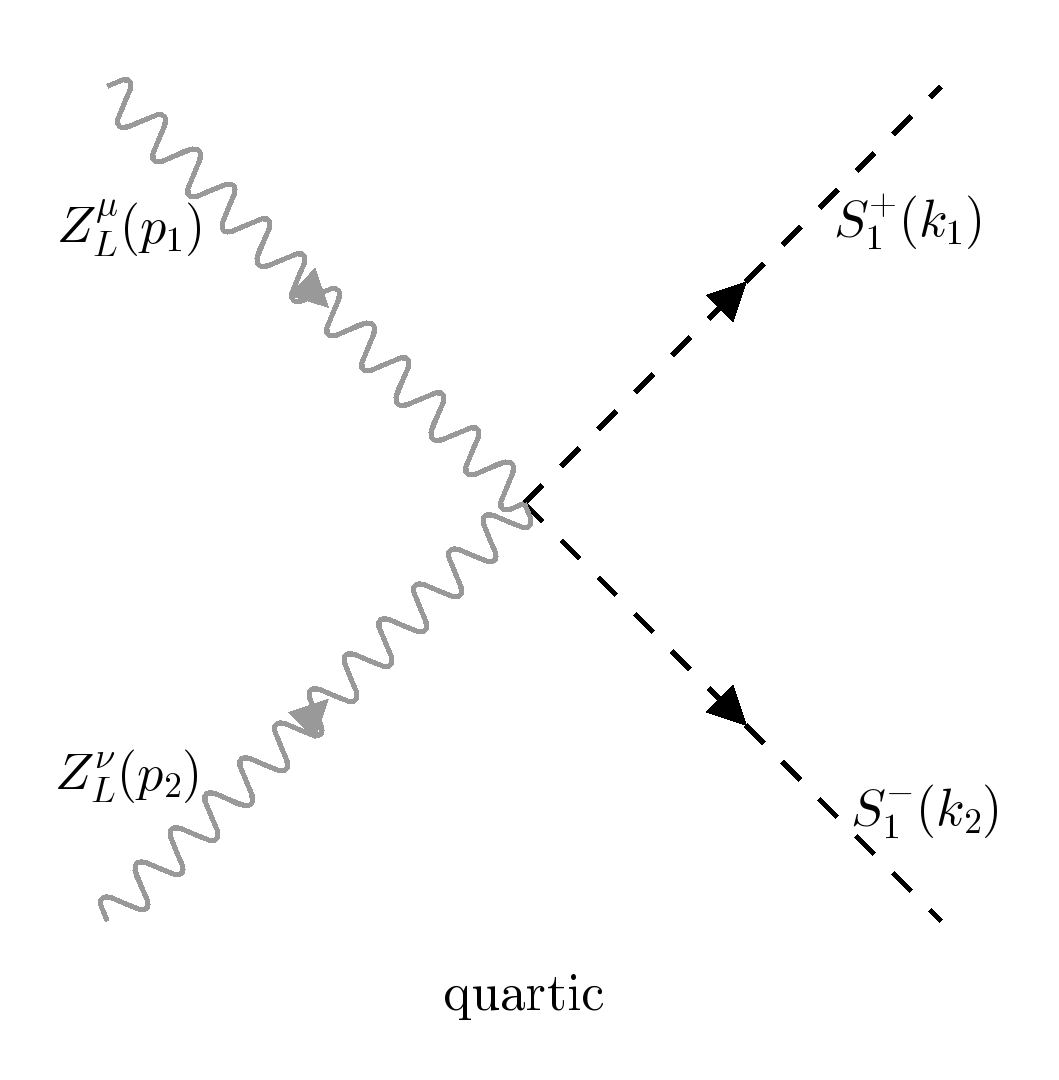}
	\caption{Feynman diagrams for $Z_L Z_L \to S_1^+ S_1^-$.}
	\label{f:FD1}
\end{figure}
The vertex factor for the interaction $Z_\mu S_1^+(p) S_2^-(p^\prime)$ is written as $i g_{z s_1 s_2} (p - p^\prime)_\mu $ where, $p$ and $p^\prime$ 
are the momenta of the incoming charged scalars. 
Considering the momentum assignment of the initial and final states particles as in Eq.~(\ref{e:P1}), we can write
\begin{eqnarray}
	p_1 + p_2 = k_1 + k_2 \,.
	\label{eq:momcons}
\end{eqnarray}
Now, the Feynman amplitude for the $t$-channel diagram is given by,
\begin{eqnarray}
	i{\cal M}_t &=& (i g_{z s_1 s_2})^2 \left[-(p_1 - k_1)+k_1\right]_\mu \frac{i}{t - m_{C_2}^2} \left[-k_2 - \left(p_1 - k_1\right)\right]_\nu \epsilon^\mu(p_1) \epsilon^\nu(p_2) \,, \nonumber \\
	&=& \frac{i g_{z s_1 s_2}^2}{t - m_{C_2}^2} \left(p_1 - 2k_1\right)_\mu \epsilon^\mu(p_1) \left(p_2 - 2 k_2 \right)_\nu \epsilon^\nu(p_2) \,,
	\label{eq:mattchan}
\end{eqnarray}
where we used Eq.~(\ref{eq:momcons}) in the last step. In a similar manner, we write down the 
matrix element for the $u$-channel diagram as,
\begin{eqnarray}
	i{\cal M}_u &=& (i g_{z s_1 s_2})^2 \left[-k_2 - \left(k_2 - p_1\right)\right]_\mu \frac{i}{u - m_{C_2}^2} \left[k_1 - \left(k_2 - p_1\right)\right]_\nu \epsilon^\mu(p_1) \epsilon^\nu(p_2) \,, \nonumber \\
	&=& \frac{i g_{z s_1 s_2}^2}{u - m_{C_2}^2} \left(p_1 - 2k_2\right)_\mu \epsilon^\mu(p_1) \left(p_2 - 2 k_1 \right)_\nu \epsilon^\nu(p_2) \,.
	\label{eq:matuchan}
\end{eqnarray}
Next, we express the longitudinal polarization vector for the $Z$-boson as $\epsilon_L^\mu(p) \equiv {\epsilon^\mu(p)}/{M_Z}$
with the understanding that $\epsilon^\mu(p)\epsilon_\mu(p)= -M_Z^2$ and $p_\mu \epsilon^\mu(p)= 0$. 
The kinematics for the process in the CM frame has been schematically depicted in fig.~\ref{f:kine}.
Following this, we may write
\begin{subequations}
	\label{eq:pola}
	\begin{eqnarray}
	&&	k_1 \cdot \epsilon_L(p_1) = \frac{E}{M_Z}(p - k\cos\theta) = k_2 \cdot \epsilon_L(p_2)\,; \\
	&& k_2 \cdot \epsilon_L(p_1) = \frac{E}{M_Z}(p + k\cos\theta) = k_1 \cdot \epsilon_L(p_2)\,.
	\end{eqnarray}
\end{subequations}
Using the relations given in Eq.~(\ref{eq:pola}), we now rewrite the matrix elements as,
\begin{subequations}
	\label{eq:mat2}
	\begin{eqnarray}
	{\cal M}_t &=& 4  g_{z s_1 s_2}^2\frac{E^2}{M_Z^2} \frac{(p -k \cos\theta)^2}{t - m_{C_2}^2} \,,  \\
	{\cal M}_u &=& 4 g_{z s_1 s_2}^2\frac{E^2}{M_Z^2} \frac{(p + k \cos\theta)^2}{u - m_{C_2}^2} \,.
	\end{eqnarray}
\end{subequations}
Now, using $t = (p_1 - k_1)^2$ and $u = (p_2 - k_1)^2$ in combination with Eq.~(\ref{eq:momcons}), we can show that
\begin{subequations}
\label{eq:delm}
	\begin{eqnarray}
	\left(p - k \cos \theta \right) &=& \frac{\Delta m^2 - t}{2 p} \,, \\
	\left(p + k \cos \theta \right) &=& \frac{\Delta m^2 - u}{2 p} \,,
	\end{eqnarray}
\end{subequations}
where, we have defined $\Delta m^2 = m_{C_1}^2 - M_Z^2$.
Thus, substituting Eq.~(\ref{eq:delm}) into Eqs.~(\ref{eq:mat2}), we find,
\begin{subequations}
	\begin{eqnarray}
	{\cal M}_t &=& \frac{g_{z s_1 s_2}^2 E^2}{p^2 M_Z^2} \left(\Delta m^2 - t\right)^2 \frac{1}{t}\left(1 - \frac{m_{C_2}^2}{t}\right)^{-1} \,, \\	
	\Rightarrow	{\cal M}_t	&\simeq& \frac{g_{z s_1 s_2}^2 E^2}{p^2 M_Z^2} \left(t - 2 \Delta m^2 + m_{C_2}^2 \right) \,,	\\
	{\rm and,~ similarly} \quad {\cal M}_u	&\simeq& \frac{g_{z s_1 s_2}^2 E^2}{p^2 M_Z^2} \left(u - 2 \Delta m^2 + m_{C_2}^2 \right) \,,
	\end{eqnarray}
\end{subequations}
where, we have neglected the terms ${\cal O}(M^2/E^2)$ in the high-energy limit. 
Furthermore, using the identity
\begin{eqnarray}
	\frac{E^2}{p^2} = E^2 (E^2 - M_Z^2)^{-1} \simeq \left(1 + \frac{M_Z^2}{E^2} \right) \,,
\end{eqnarray}
we reduce the above amplitudes into,
\begin{subequations}
	\begin{eqnarray}
	{\cal M}_t &=&  \frac{g_{z s_1 s_2}^2}{M_Z^2} t + g_{z s_1 s_2}^2 \left[\frac{t}{E^2} - \frac{\Lambda^2}{M_Z^2} \right] + {\cal O}\left(\frac{M^2}{E^2}\right) \,,  \\
	{\rm and,}\quad		{\cal M}_u &=&  \frac{g_{z s_1 s_2}^2}{M_Z^2} u + g_{z s_1 s_2}^2 \left[\frac{u}{E^2} - \frac{\Lambda^2}{M_Z^2} \right] + {\cal O}\left(\frac{M^2}{E^2}\right) \,,
	\end{eqnarray}
\end{subequations}
where we defined $\Lambda^2 = (2 \Delta m^2 - m_{C_2}^2)$.
Next, the Feynamn amplitude for the quartic diagram is,
\begin{eqnarray}
	i{\cal M}_Q &=& 2 i(g_{z s_1 s_2})^2 \epsilon^\mu_L(p_1) \epsilon_{\mu L}(p_2) \,. 
	\label{eq:matquart}
\end{eqnarray}
Replacing the longitudinal polarization vector for the $Z$-boson as $\epsilon_L^\mu(p) \equiv {\epsilon^\mu(p)}/{M_Z}$
we get
	\begin{eqnarray}
		&&	\epsilon_L(p_1)  \cdot \epsilon_L(p_2) =  \frac{1}{M_Z^2}\epsilon(p_1)  \cdot \epsilon(p_2)  = \left(\frac{2E^2}{M_Z^2} - 1\right) \,.
	\end{eqnarray}
Thus the total amplitude will be given by,
{\small
\begin{eqnarray}
{\cal M}_{Z_L Z_L \to S_1^+ S_1^-} &=& ({\cal M}_t + {\cal M}_u + {\cal M}_Q)
\nonumber \\ 
&=& \frac{g_{z s_1 s_2}^2}{M_Z^2} (t + u) + g_{z s_1 s_2}^2 \left[\frac{(t+u)}{E^2} - \frac{2\Lambda^2}{M_Z^2} \right] + 2g_{z s_1 s_2}^2 \left(\frac{2E^2}{M_Z^2} - 1\right)  +{\cal O}\left(\frac{M^2}{E^2}\right) \,.
\end{eqnarray}
}
Substituting $t+u = 2(M_Z^2 + m_{C_1}^2) - 4 E^2 $ we obtain,
{\small
\begin{subequations}
	\begin{eqnarray}
	{\cal M}_{Z_L Z_L \to S_1^+ S_1^-} &=& -\frac{4 g_{z s_1 s_2}^2}{M_Z^2} E^2 + \frac{g_{z s_1 s_2}^2}{M_Z^2} \left[2(M_Z^2 + m_{C_1}^2) - 2\Lambda^2 - 4 M_Z^2 \right]  \nonumber \\
	&& + 2g_{z s_1 s_2}^2 \left(\frac{2E^2}{M_Z^2} - 1\right) +{\cal O}\left(\frac{M^2}{E^2}\right) \,.
 	\end{eqnarray}
Now we can use the definition of $\Lambda$ to write
	\begin{eqnarray} 
	\Rightarrow {\cal M}_{Z_L Z_L \to S_1^+ S_1^-}   &\approx&  -\frac{4 g_{z s_1 s_2}^2}{M_Z^2} E^2 + \frac{2 g_{z s_1 s_2}^2}{M_Z^2} \left[ M_Z^2 + \left( m_{C_1}^2 - m_{C_2}^2 \right) \right] + 2g_{z s_1 s_2}^2 \left(\frac{2E^2}{M_Z^2} - 1\right) \,,  \\
	\Rightarrow {\cal M}_{Z_L Z_L \to S_1^+ S_1^-}  &\approx&  \frac{2 g_{z s_1 s_2}^2}{M_Z^2}\left( m_{C_1}^2 - m_{C_2}^2 \right) \,.
	\label{eq:mtot}
	\end{eqnarray}
\end{subequations}
}
%
It is quite interesting to note that the energy growths of ${\cal O}(E^2)$
arising from the $t$ and $u$ channel diagrams get exactly canceled by the
quartic diagram as expected in spontaneously broken gauge theories.

Next we consider the process
\begin{eqnarray}
	\centering
	\Huge
	Z_L(p_1) + Z_L(p_2) \to S_1^+(k_1) + S_2^-(k_2) \,.
\end{eqnarray}
Since we are assuming the presence of off-diagonal couplings only, with the Higgs and
the $Z$-boson, the above process can only proceed via the $s$-channel Higgs exchange.
The corresponding amplitude will be given by
\begin{subequations}
	\begin{eqnarray}
	i{\cal M}_{Z_L Z_L \to S_1^+ S_2^-} &=& \left(\frac{i g M_Z}{c_w}\right) \frac{i}{s - m_h^2} \left(i \lambda_{hs_1s_2} M_W \right) \epsilon_L^\mu(p_1)\epsilon_{L\mu}(p_2) \,, \\
	\Rightarrow {\cal M}_{Z_L Z_L \to S_1^+ S_2^-}  &=& \left(- g M_Z^2 \lambda_{hs_1s_2}\right) \frac{1}{s - m_h^2} \frac{\left(p^2 + E^2\right)}{M_Z^2}  \,,
	\end{eqnarray}
\end{subequations}
where we used $p_1^\mu = (E, p\hat{z})$ and $p_2^\mu = (E, -p\hat{z})$ in the CM frame. 
Furthermore, using $p^2 + E^2 = s/2 - M_Z^2$ we get,
\begin{eqnarray}
	{\cal M}_{Z_L Z_L \to S_1^+ S_2^-} \approx -\frac{1}{2}g \lambda_{hs_1s_2} + {\cal O}\left(\frac{M^2}{E^2}\right) \,. 
	\label{eq:math}
\end{eqnarray}

\begin{figure}[htbp!]
	\centering
	\includegraphics[scale=0.115]{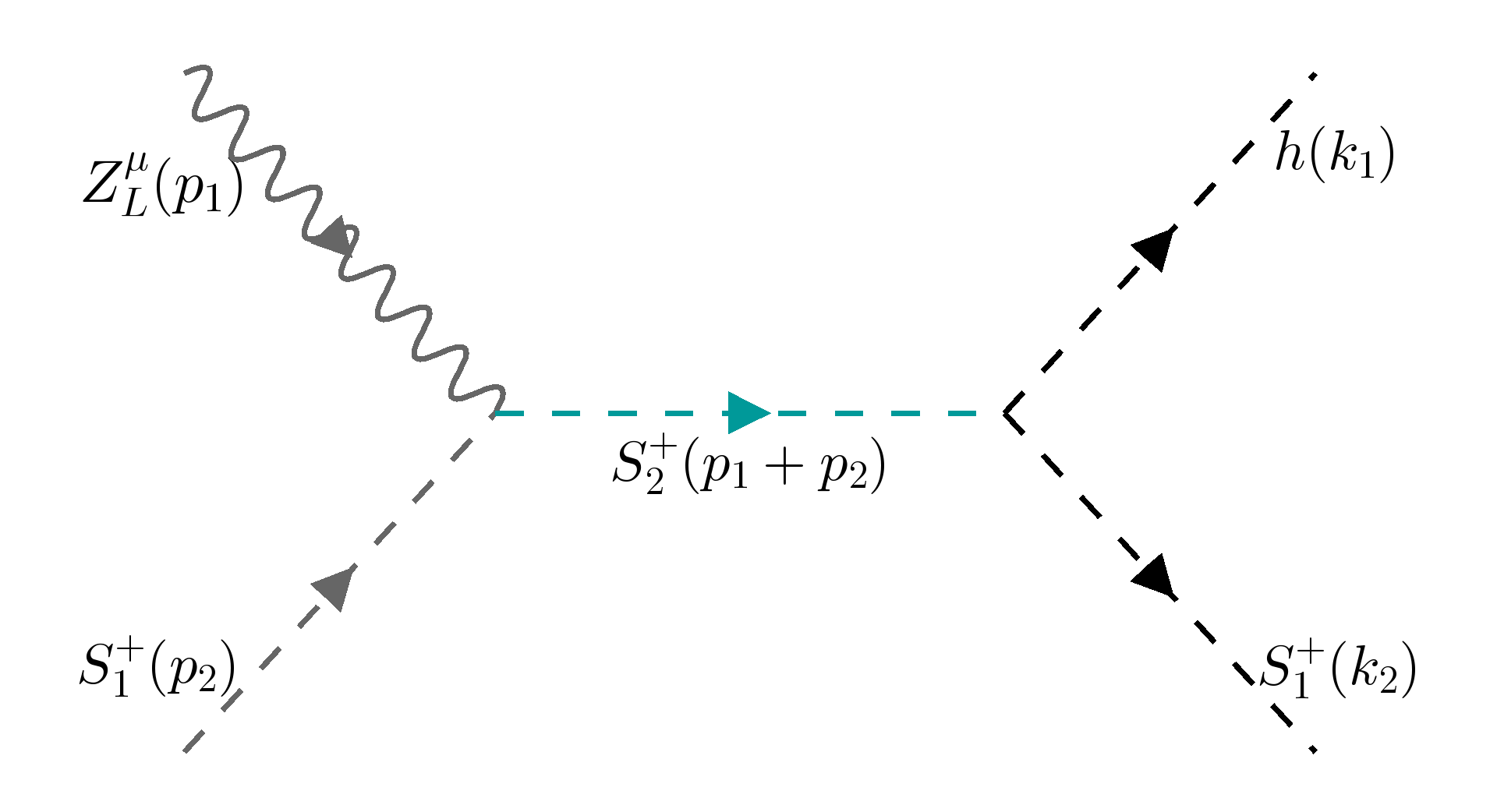} ~~
		\includegraphics[scale=0.125]{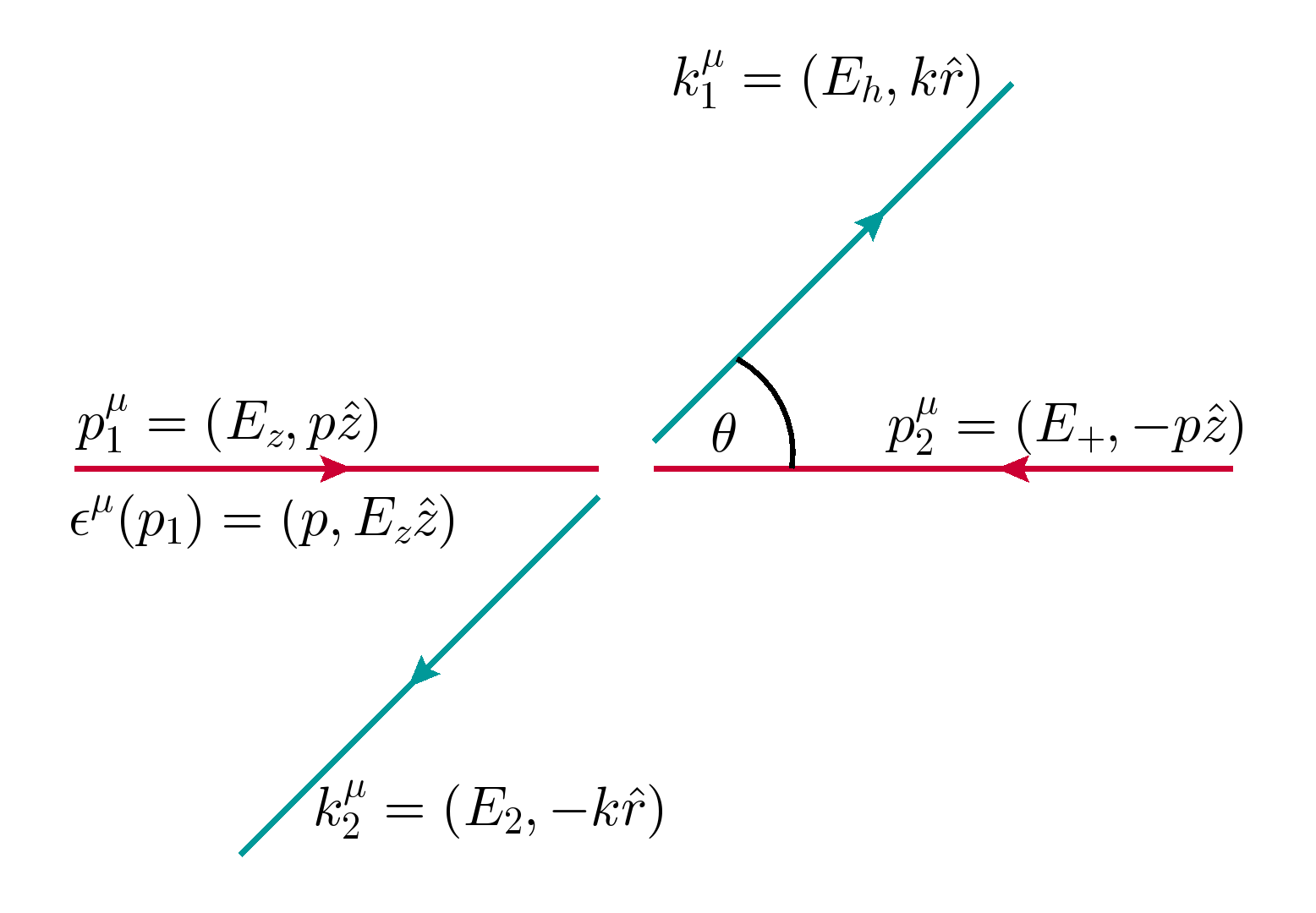}
	\caption{Feynman diagram for the process appearing in \Eqn{e:p3} and the corresponding kinematics
		in the CM frame.}
	\label{f:FD3}
\end{figure}

Finally, we consider the process
\begin{eqnarray}
	\centering
	\Huge
	Z_L(p_1) + S^+_1(p_2) \to h (k_1) + S_1^+(k_2) \,.
	\label{e:p3}
\end{eqnarray}
The Feynman diagram along with the kinematics in the CM frame have been depicted in Fig.~\ref{f:FD3}.
%
The amplitude for the process may be written as,
\begin{subequations}
	\begin{eqnarray}
	&&	i{\cal M}_{Z_L S_1^+ \to h S_1^+} = ig_{zs_1s_2} \left\{p_2 +(p_1 + p_2)\right\}_\mu \frac{i}{s - m_{C_2}^2} (i\lambda_{hs_1s_2}M_W)
	\epsilon_{L}^\mu(p_1) \,, \\
	\Rightarrow && {\cal M}_{Z_L S_1^+ \to h S_1^+} = -  g_{zs_1s_2}\lambda_{hs_1s_2} \frac{M_W}{M_Z} \frac{1}{s - m_{C_2}^2} \left(2 p_2 \cdot \epsilon(p_1) \right) \,.
	\label{e:MZShS}
	\end{eqnarray}
\end{subequations}
Now, following the kinematics in Fig.~\ref{f:FD3} we have the following relations,
\begin{subequations}
	\begin{eqnarray}
	&&	p_2 \cdot \epsilon (p_1)=  (E_+ p + E_z p) = p(E_+ + E_z) \equiv p \sqrt{s} \,, \\
	{\rm and,} && p^2 = E_z^2 - M_Z^2 = E_+^2 - m_{C_1}^2 \,, \\
	\Rightarrow &&  M_Z^2 - m_{C_1}^2 = E_z^2 - E_+^2 \,, \\
	\Rightarrow && (E_z - E_+) = \frac{M_Z^2 - m_{C_1}^2}{\sqrt{s}} \,.
	\end{eqnarray}
\end{subequations}
Alternatively, one can also write,
\begin{subequations}
	\begin{eqnarray}	
	2p^2 &=& E_z^2 + E_+^2 - (M_Z^2 + m_{C_1}^2) \,, \nonumber\\
	&=& \frac{1}{2}\left[(E_+ + E_z)^2 + (E_+ - E_z)^2\right] - (M_Z^2 + m_{C_1}^2) \,, \nonumber \\
	&=& \frac{1}{2}\left[ s + \frac{(M_Z^2 - m_{C_1}^2)^2}{s} - 2(M_Z^2 + m_{C_1}^2)\right] \,, \\
	\Rightarrow	p^2 &=& \frac{1}{4}\left[ s + \frac{(M_Z^2 - m_{C_1}^2)^2}{s} - 2(M_Z^2 + m_{C_1}^2)\right] \,, \\
	\Rightarrow p &=& 	\frac{\sqrt{s}}{2} \left[ 1 - 2\frac{(M_Z^2 + m_{C_1}^2)}{s} + \frac{(M_Z^2 - m_{C_1}^2)^2}{s^2} \right]^{\frac{1}{2}} \,, \\ 
	p &\approx& \frac{\sqrt{s}}{2} \left[ 1 - \frac{(M_Z^2 + m_{C_1}^2)}{s} + {\cal O}\left(\frac{M^4}{s^2}\right) \right]  \,.
	\end{eqnarray}
\end{subequations}
Thus, one may write
\begin{eqnarray}
p_2\cdot\epsilon(p_1) &=& \frac{s}{2} \left[ 1 - \frac{(M_Z^2 + m_{C_1}^2)}{s} + {\cal O}\left(\frac{M^4}{s^2}\right) \right] \,.
\end{eqnarray}
Using this in \Eqn{e:MZShS}, the final expression for the amplitude, in the high-energy limit, can be written as
\begin{eqnarray}
	{\cal M}_{Z_L S_1^+ \to h S_1^+} \approx - g_{zs_1s_2} \lambda_{hs_1s_2} \frac{M_W}{M_Z} + {\cal O}\left(\frac{M^2}{s}\right) \,.
\end{eqnarray}


\bibliographystyle{JHEP}
\bibliography{ref}

\end{document}